\documentclass[aps,floatfix,showpacs,prb]{revtex4}
\usepackage{graphicx,amsfonts,amssymb,enumerate,color}
\usepackage{natbib}
\newcommand{\be}{\begin{equation}}
\newcommand{\ee}{\end{equation}}
\newcommand{\bea}{\begin{eqnarray}}
\newcommand{\eea}{\end{eqnarray}}
\newcommand{\beb}{\begin{eqnarray*}}
\newcommand{\eeb}{\end{eqnarray*}}

\newcommand{\phrb}[3]{Phys. Rev. B{\bf #1}, #2 (#3).}
\newcommand{\phrl}[3]{Phys. Rev. Lett. {\bf #1}, #2 (#3).}
\newcommand*{\vcenteredhbox}[1]{\begin{tabular}{@{}c@{}}#1\end{tabular}}
\begin{document}
\title{Competing Laughlin state and Wigner crystal in bilayer graphene}

\author{Ngoc Duc Le}
\author{Thierry Jolicoeur}
\affiliation{Universit\'e Paris-Saclay, CNRS, CEA, Institut de Physique Th\'eorique, France}

\date{October 13th, 2022}

\begin{abstract}
We study the fractional quantum Hall effect in the central Landau level
of bilayer graphene. By tuning the external applied magnetic field and
the electric bias between the two layers one can access a regime where
 there is a degeneracy between Landau levels with orbital characters
corresponding to N=0 and N=1 Galilean Landau levels.
While the Laughlin state is generically the ground state for filling $\nu=1/3$
we find that it can be destroyed and replaced by a Wigner crystal
at the same filling factor by tuning the bias and applied field.
This competition does not take place at $\nu=2/3$ where the
incompressible ground state remains stable. The possibility of electrically
inducing the Wigner crystal state opens a new range of studies of this
state of matter.
\end{abstract}
\pacs{73.43.-f, 73.22.Pr, 73.20.-r}
\maketitle

\section{Introduction}

Graphene-based samples have revealed an abundance of correlated phases
in the quantum Hall 
regime~\cite{Du,Bolotin,Ghahari,Dean-MLG1,Young-MLG2,Feldman-MLG3,Feldman-MLG4,Hunt,
Young-MLG4,Amet-MLG5,Zibrov-MLG6,Polshyn-MLG7,Chen-MLG8,Zeng-MLG9,Zhou-MLG10,Yang-MLG11}. 
Odd and even-denominator
fractional quantum Hall states are observed in monolayer graphene as well as in
bilayer graphene, fractional Chern insulators have been realized in samples
involving graphene-hexagonal boron nitride structures.
Also of interest is the field-induced excitonic condensate in double-bilayer 
samples. It has been pointed out that bilayer graphene (BLG) systems have a convenient parameter which can be tuned experimentally~: the interlayer electric bias in addition to
electronic density and external applied magnetic field.
The central Landau level of BLG has a nearly eightfold degeneracy due
to the combination of several quantum numbers~: the ordinary spin, the valley degree of 
freedom and also an orbital degeneracy. The pattern of ordering in these levels
is then very rich and complex. It has been shown that there are tunable phase transitions
in fractional quantum Hall states~\cite{Apalkov2010}. The electric bias directly
controls the splitting between orbital levels and the Coulomb interaction between electrons
is also impacted by the value of the external applied magnetic field as well as the bias.
Detailed
investigations of the integer quantum Hall states have been performed
on these systems~\cite{Zibrov2017} and have shown that an appropriate tight-binding model
can capture the level ordering. Recent advances have led to the observation of many fractional states and transitions between them. This means that we have at our disposal a physical system where we can tune
parameters affecting the fractional quantum Hall 
physics~\cite{Maher-BLG1,Maher-BLG2,Kou-BLG3,Li-BLG4,Pientka-BLG5,Hunt-BLG6,Fu-BLG7,Seiler-BLG8,Huang2022}.

In the case of two-dimensional electron systems in GaAs it is known that
there is a competition between incompressible electron liquids and crystal of electrons, the so-called 
Wigner crystal. For filling factor $\nu=1/3$ of the lowest Landau level, the ground state
of the electronic system with Coulomb interactions is an
incompressible liquid whose properties are well described by the Laughlin wavefunction~\cite{Laughlin} and 
it is only for small filling factors that the ground state is a crystal state~\cite{LeGlatt}.
Deciding the precise boundary between these phases has proven a difficult issue~\cite{Zuo2020}.
The crystal state appears as insulating state with a diverging longitudinal resistance when
decreasing the temperature. There are experimental evidences for reentrance of the Wigner crystal
when one  decreases the filling factor. Studies of the crystal state are difficult due to the large values
of the magnetic field needed to destroy the fractional quantum Hall liquids. The crystal state is not
the only competitor with the liquid states. In higher Landau levels it is known that the electronic system
may also form the so-called stripe or bubble phases. As the Wigner crystal such states break translation
symmetry
and it is believed that they are in a distinct state of matter without topological order.
Their experimental signature is an insulating behavior with additional anisotropic properties.
We note that in two-dimensional GaAs electron or
hole systems~\cite{Shayegan98,Santos92,Santos92II,Manoharan96,MengMa2020} 
there is a rich competition between several many-body ground states and that
the Wigner crystal can be stabilized at or close to filling 1/3 by tuning gate potentials.

Graphene systems offer yet another arena for the study of such competing phases, in particular the bilayer graphene
due to its tunability. It is also known that mixing with higher Landau levels can bias the competition
towards the Wigner crystal state.
Tuning the BLG system to obtain 
degeneracy of Landau levels with N=0 and N=1 character can be viewed as an extreme example of Landau level
mixing albeit with absence of levels with $N>1$. So it is plausible that the competition between the Laughlin
state and the Wigner crystal can be tuned.

In this paper we investigate the incompressible quantum Hall states that occur
for filling factor $\nu=1/3$ and $\nu=2/3$ when the system is fully valley as well as spin polarized 
in the bilayer graphene system. 
The interesting physics now emerges from the crossing of levels
with orbital character N=0 and N=1. This should happen in the central octet of Landau levels
for fillings close to $\nu=-3$ according to the present knowledge of level orderings~\cite{Hunt-BLG6}.
Electrons form then an effective two-component system
with a tunable anisotropic interaction given by the projection of the Coulomb potential
into this two-component subspace. We use exact diagonalization technique in the torus geometry which 
is well suited to study the competition between incompressible quantum Hall phases and the Wigner
crystal~\cite{HRY1999,HRY2000,YHR2001,Wang2008}. 

For the filling factor $\nu=1/3$ we find that the Laughlin-like ground state is stable in a wide region
of the phase diagram of the BLG system. However there is a region enclosing the degeneracy point
where it is no longer the ground state. We observe the appearance of a set of quasi-degenerate ground states 
that form a lattice in reciprocal space as expected for spontaneous breakdown of translational symmetry.
The crystalline correlations are also revealed by computing an appropriate pair correlation function.
The crystal state is fully polarized in the $N=0$ orbitals since the exchange energy is larger than in 
the $N=1$ case. It appears close to the boundary of a polarization transition. The transition to the Laughlin
state is either second order or weakly first order since the the ground state manifold includes a zero-momentum
state that smooth deforms into the Laughlin state.

In the case of the fraction $\nu=2/3$ we find no such crystal state but only a smooth crossover
between the fully spin polarized state which is the particle-hole transform of the $\nu=1/3$ state
and the singlet state.

In section \ref{QHEBLG} we describe the consequences of the band structure of bilayer graphene
onto the Landau levels and discuss the effective Hamiltonian we use.
Section \ref{Laughlin} discusses our findings about the Laughlin state and the Wigner crystal.
Section \ref{TwoThirds-section} is devoted to the fraction $\nu=2/3$ where incompressible liquids
are always stable.
Finally section \ref{concluding} presents our conclusions.

\section{quantum Hall effect in bilayer graphene}
\label{QHEBLG}
Bilayer graphene under a magnetic field has a peculiar Landau level structure.
Notably the zeroth level has an approximate eightfold degeneracy.
These eight levels have valley twofold indices $\tau=K,K^\prime$,
spin $\sigma=\uparrow,\downarrow$ and also an orbital index $N=0,1$
which is unique to the bilayer system. Coulomb interactions are isotropic in spin and valley space
to a very good approximation but there is no good symmetry between the two orbital states.
We denote the wavefunctions~\cite{McCann2012,Snizhko2012} of these states as 
$\psi_{N\sigma\tau}$ with $N=0,1$.
The $\psi_{1\sigma\tau}$ wavefunctions have weight concentrated onto the $n=0$ and $n=1$ cyclotron
Galilean orbitals that we denote $\phi_{0,1}$~:
$\psi_{1K\uparrow}=(\sqrt{1-\gamma}\phi_1,0,\gamma\phi_0,0)$ and
$\psi_{1K^\prime\uparrow}=(0,\sqrt{1-\gamma}\phi_1,0,\gamma\phi_0)$ where the four components
denote amplitudes on the sites $A,B^\prime,A^\prime,B$ where the sites are defined in 
ref.(\onlinecite{McCann2012}). In what follows $N$ stands for BLG orbitals while $n$ refers to Galilean
orbitals deriving from electrons with a parabolic dispersion relation.
Similarly we have $\psi_{0K\uparrow}=(\phi_0,0,0,0)$ and
$\psi_{0K^\prime\uparrow}=(0,\phi_0,0,0)$. The $\gamma$ parameter can be adjusted
by the external magnetic field and has an almost linear variation
from $\gamma=0.01$ for $B=2T$ to $\gamma=0.35$ at $B=45T$ according to Ref.(\onlinecite{Zibrov2017}).
This parameter also varies with the electric bias $u$ between the layers.

In this work we concentrate on the case of spin and valley polarized configurations
but we consider possible degeneracy of $N=0$ and $N=1$ states. We thus project the Coulomb interaction
onto this two-component subspace. So we make the approximation of neglecting the mixing to
higher Landau levels, necessary to reduce the Fock space dimension to
a manageable size.
The second-quantized Hamiltonian can now be written as~:
\be
\mathcal{H}=\frac{1}{2A}\sum_{q\{N_i\}}
{\tilde V}(q) F_{N_1 N_2}(q)F_{N_3 N_4}(-q)
\colon \rho^\dag_{N_1N_2}(q) \rho_{N_3N_4}(q)\colon\quad
+\quad \Delta_{10}{\hat N}_0
\label{HamiltonianBLG}
\ee
The operators $\rho_{nm}(q)$ are projected density operators 
with the following form factors~:
\be
F_{00}=\textrm{e}^{-q^2/4},\quad F_{11}=((1-\gamma)L_1(q^2)+\gamma L_0(q^2))\textrm{e}^{-q^2/4},\quad
F_{10}=F_{01}^{*}=\sqrt{1-\gamma}\,\,\frac{q_y-iq_x}{\sqrt{2}}\textrm{e}^{-q^2/4}
\label{FormFactors}
\ee
Here we use the Coulomb potential ${\tilde V}(q)=2\pi e^2/\epsilon |q|$, $A$
is the area of the system. $L_0$ and $L_1$ are Legendre polynomials.
In BLG systems the screening effects may modify the Coulomb potential.
We have used the modification proposed in Ref.(\onlinecite{PapicAbanin})
and checked that our results do not depend sensitively upon screening effects
as was observed previously~\cite{JTS}. So the results presented in this paper
are for the bare Coulomb potential.
We measure all lengths in units of the magnetic length
$\ell=\sqrt{\hbar/eB}$ and energies are measured via the Coulomb scale
$e^2/(\epsilon\ell)$.
We explore the FQHE phase diagram as a function of the $\gamma$ parameter which 
has a range $[0,1]$ and also as a function of the level splitting $\Delta_{10}$.
While strictly speaking the range of this parameter is unbounded all interesting
variations occur if we vary it in the the interval $\Delta_{10}=+1\dots -1$
in units of the Coulomb energy scale.
In bilayer graphene the splitting $\Delta_{10}$ is controlled mostly by the interlayer bias $u$.
The precise relation has been explored in some details~\cite{Zibrov2017}. At zero bias $u=0$ there is
nevertheless a nonzero splitting of $N=0$ and $N=1$ states due to the Lamb shift-like effect
of the Fermi sea of all filled levels below the central octet. Achieving degeneracy $\Delta_{10}=0$
requires tuning $u$ to negative values in the conventions of ref.(\onlinecite{Zibrov2017}).
In Fig.(\ref{PhaseDiag}) we have plotted three lines giving the values of $\gamma$ and $\Delta_{10}$
when we vary the magnetic field at fixed interlayer bias $u=+80,0,-80$~meV.

Several limiting cases can be reached as a function of these parameters.
When $\gamma=1$ the Coulomb interaction is identical to that of the lowest Landau
level
of nonrelativistic electrons irrespective of the orbital index since then we have 
$F_{00}=F_{11}$
and there are no Coulomb couplings between the orbitals $F_{10}=F_{01}=0$. So 
there is complete $SU(2)$
symmetry in the $n=0,1$ space and the problem is formally equivalent to
a Galilean lowest Landau level system with zero Zeeman energy if $\Delta_{10}=0$.
Varying  $\Delta_{10}$ at $\gamma=1$ is exactly like applying a Zeeman energy
on the $SU(2)$ symmetric Coulomb interacting lowest Landau level.
So we recover the physics of the FQHE in the lowest Landau level with two components.
Many fractions are known to be present in this limiting case
and their properties are mostly understood~\cite{JainBook} at least for
$\nu=1/3$ and $\nu=2/3$ most prominent cases.

If now we set $\gamma$ to zero we have two orbitals with the form factors
of the lowest $n=0$ and second $n=1$ Landau levels whose degeneracy is controlled
again
by $\Delta_{10}$. For $\Delta_{10}$ large and negative we are back to a single polarized
lowest Landau level while for $\Delta_{10}$ large and positive the physics is that
of a single polarized \textit{second} Landau level. If $\Delta_{10}=0$ we have 
maximal mixing
between the two degenerate $n=0$ and $n=1$ orbitals, a situation unique to bilayer
graphene. This is a situation which is reminiscent of strong Landau level mixing.
Indeed when the cyclotron energy is not very large with respect to the Coulomb energy scale
virtual transitions towards empty levels will modify the interactions between electrons
residing in the LLL. This effect in general destabilizes the FQHE states~\cite{Zhao} and favors
the appearance of the Wigner crystal state as is the case for example in two-dimensional hole
systems~\cite{MengMa2020}
So the two-component system at $\gamma=0$
can possibly share some of the physics of Landau level mixing.

When $\Delta_{10}$ is very large and negative all electrons stay in the $N=0$ orbitals
and the presence of the $N=1$ orbitals is manifest only through
 virtual interactions. This effect vanishes for very large
 $\Delta_{10}$ which means that the $\gamma$ parameter does no longer play any role.
 One observes then the physics of fully polarized electrons with the standard bare Coulomb interactions
 in the lowest Galilean Landau level.
 For $\Delta_{10}$ is very large and positive all electrons are now in the $N=1$ levels
 with a $\gamma$-dependent interaction as given by $F_{11}$. This is a fully polarized situation
 which interpolates between the lowest Landau level interaction for $\gamma=1$ and 
 the $n=1$ Landau level interaction for $\gamma=0$. This last case is also known to
 harbor quantum Hall liquids but with the added complication of competing phases.
 We note that for $\gamma=1/2$ the interaction is exactly the one which is relevant for the $N=1$
 Landau level of monolayer graphene.

To make progress we perform exact diagonalizations of the many-body problem
defined by Eq.(\ref{HamiltonianBLG}) in a $L_x\times L_y$ rectangular unit
cell with periodic bounary conditions. The total number of flux quanta through this system 
is quantized to an integer value
$N_\phi=L_xL_y/2\pi$. This unit cell has an adjustable parameter called the aspect
ratio $AR=L_x/L_y$.
The periodic boundary conditions are imposed on this unit cell
and we construct the many-body translation operators that are used to classify
the eigenstates~\cite{HRY1999,Haldane87}. The corresponding eigenvalues are $K_x=2\pi s/L_x$, $K_y=2\pi t/L_y$ 
where $s$ and $t$
are integers varying in the interval $0,..,N$ where $N$ is the greatest common denominator
of $N_e$ and $N_\phi$. The origin of the quantum numbers $s,t$ is not always zero.
When $pq(N_e-1)=2k+1$ for $\nu=p/q$ ($p,q$ coprime) the zero momentum state is located at 
$s_0=t_0=N/2$, otherwise
it is located at $s=t=0$. This shift is of no special physical significance.
Previous studies~\cite{JTS,Papic2011,Sodemann} for fillings $\nu=1$ and $\nu=1/2$ have revealed
a rich physics.

\section{The fate of the Laughlin state $\nu=1/3$}
\label{Laughlin}
\begin{figure}[h]
\centering
 \includegraphics[width=0.5\columnwidth]{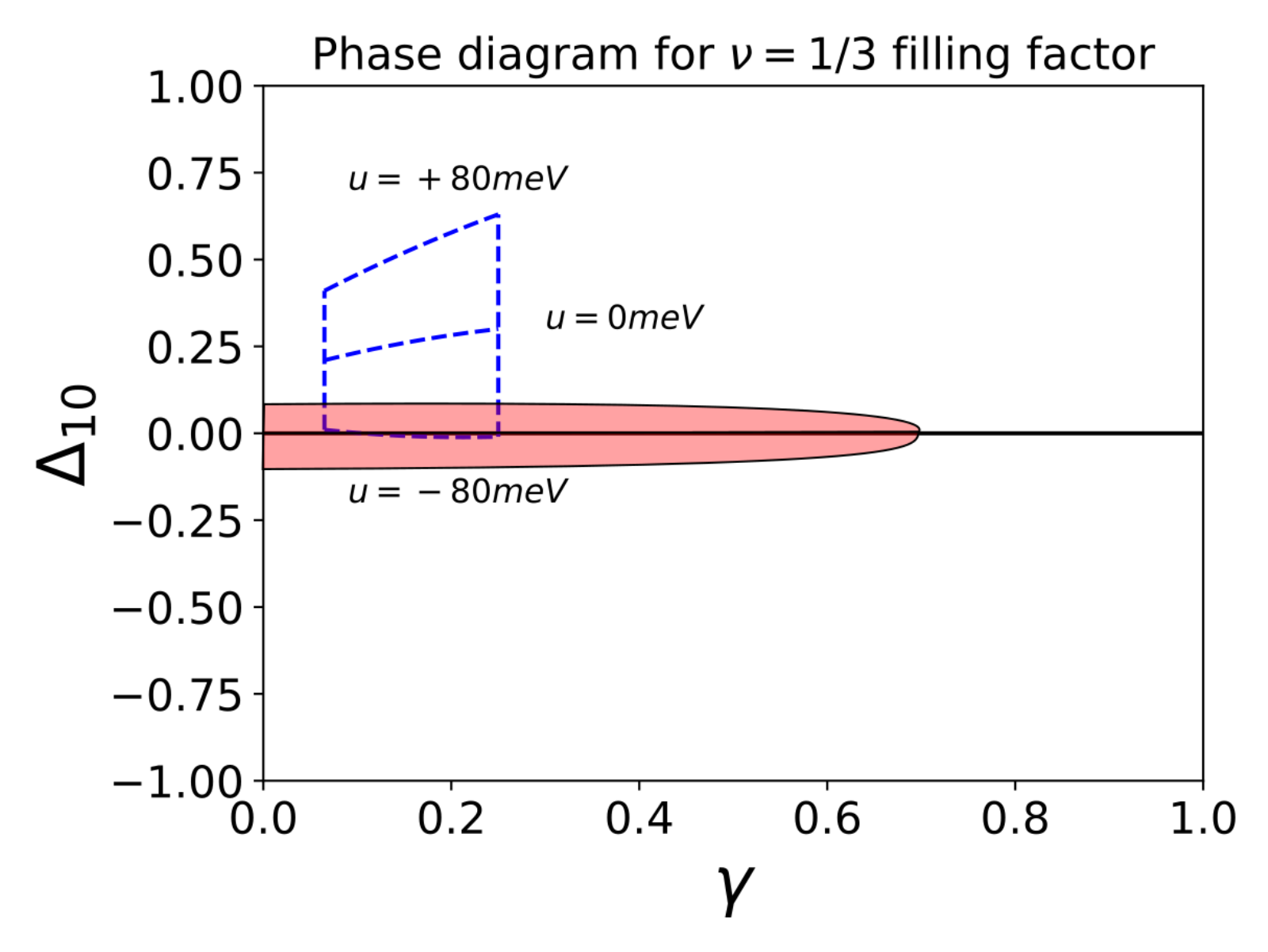}
 \caption{The phase diagram at filling factor $\nu=1/3$
 in the $\gamma=0$ and $\Delta_{10}$ plane. In the lower part corresponding to $\Delta_{10}$ negative 
 one recover the lowest Galilean Landau level physics of the Laughlin-like $\nu=1/3$ incompressible state.
 In the upper part it is the second Landau level physics which is recovered with presumably
 an incompressible state in the same universality class as the Laughlin state. The right boundary
 of the diagram with $\gamma=1$ is a spinful system of electrons with the lowest Landau level form factor.
 The non-trivial region is the neighborhood of the $\gamma=0$ and $\Delta_{10}=0$ with full mixing
 of the two orbitally distinct Landau levels of BLG. In the red region we find that the Laughlin state
 is replaced by a Wigner crystal of electrons with a triangular structure. Its extension goes up to
 $\gamma\approx 0.7$ and it is located in a narrow range of orbital splitting of at most 
 $\Delta_{10}\in [-0.08,+0.06]$ for $\gamma=0$. In devices based on BLG 
 samples~\cite{Hunt,Hunt-BLG6,Zibrov2017} typical values of
 $\gamma, \Delta_{10}$ are given by the three curves with interlayer bias $u=-80,0,+80meV$
 for magnetic field varying between $B=10T$ on the left-hand side 
 (vertical blue dotted line at $\gamma=0.06$) and $B=45T$ on the right (vertical blue line at $\gamma=0.25$).
 The Wigner crystal transition can be reached for realistic field strengths with negative enough bias.}
 \label{PhaseDiag}
\end{figure}

We now focus onto the most prominent fraction $\nu=1/3$ and study its stability
in the $\Delta_{10}-\gamma$ phase diagram. 
With the limiting values for $\gamma$ and $\Delta_{10}$ the phase diagram can be drawn in a rectangle~:
see Fig.(\ref{PhaseDiag}).


Along the $\gamma=1$ line which is the vertical rightmost boundary
the two orbitals have exactly the same form factor and there is no mixing between them since we have $F_{10}=F_{01}=0$. The external field $\Delta_{10}$ is exactly analogous to
a Zeeman field and the Coulomb interaction has the $SU(2)$ symmetry between the two components. 
At filling factor $\nu=1/3$
we should thus find the incompressible Coulomb FQHE state well approximated
by the Laughlin wavefunction~\cite{Laughlin}.
This FQHE state is fully polarized for $\Delta_{10}$ large, be it positive or negative. 
For the degeneracy point $\Delta_{10}=0$ we have an unbroken $SU(2)$
symmetry and it is known that we have then an instance of quantum Hall ferromagnetism~:
the Laughlin state is a multiplet with total spin $S_{tot}=N_e/2$ where ``spin'' refers
to the two components $N=0/1$. Adding a nonzero Zeeman-like energy will lift completely
the degeneracy of the ground state multiplet favoring the $S^z_{tot}=+N_e/2$ or $-N_e/2$ according
to the sign of $\Delta_{10}$.

Along the $\Delta_{10}=-1$ line the system becomes one-component and the Coulomb interaction
is governed by the form factor $F_{00}$ which is independent  of $\gamma$. So we recover
the lowest Landau physics of the $\nu=1/3$ Laughlin-like state.

If we now explore the $\Delta_{10}=+1$ boundary the system is also one-component but
interactions are ruled now by the form factor $F_{11}$ which is $\gamma$ dependent.
For $\gamma=1$ the interaction is exactly that of the Galilean lowest Landau level
while for $\gamma=0$ it is the interaction of the \textit{second} Landau level.
In between we observe that the value $\gamma=1/2$ gives the form factor
appropriate to the $N=1$ Landau level of monolayer graphene. We are not able to shed any new light
on the nature of the $\nu=1/3$ state in the second Landau level. It is known that finite size effects
become important and that a finite width of the electron gas is necessary to stabilize the Laughlin state.
Here we take the simple view that the Laughlin state is still a valid description of $\nu=1/3$
even in this case. As a consequence the set of boundaries (large negative $\Delta_{10}$, any $\gamma$),
(any $\Delta_{10}$, $\gamma=1$), (large positive $\Delta_{10}$, any $\gamma$) all share the same 
Laughlin-like physics differing only by polarization. So the most interesting region
is the small $\gamma$ small $\Delta_{10}$ region. We expect that varying $\Delta_{10}$
will induce a polarization transition between the two components. So
to locate phase boundaries we use two indicators~: the polarization of the ground state
and the quantum fidelity.

\subsection{Polarization transition}
\label{polar-section}
\begin{figure}[thp]
\centering
  \includegraphics[width=0.5\textwidth]{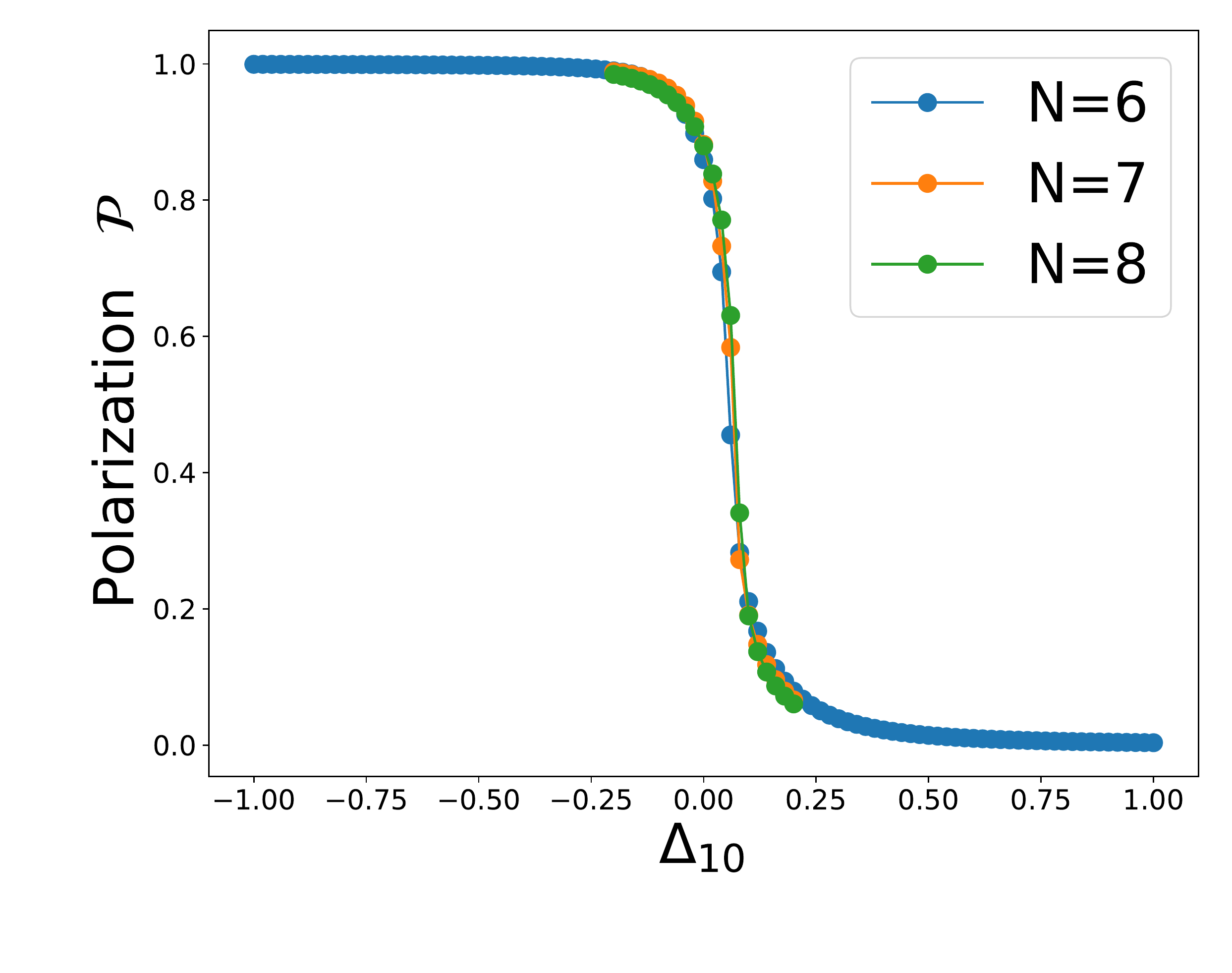}
  \caption{The polarization of the zero-momentum ground state 
 at $\nu=1/3$ for  $\gamma=0$ as a function of $\Delta_{10}$. The curve is very weakly dependent upon
 the aspect ratio of the unit cell. For large negative values all electrons prefer to occupy
 the N=0 orbital while for large positive $\Delta_{10}$ they occupy the N=1 orbitals.
 We note that the transition between these two regimes happens for a non-trivial value of the bare
 splitting $\Delta_{10}\approx 0.06$ insensitive to the system sizes we have studied. 
 Sizes are $N_e=6,7,8$ electrons}
  \label{polarplot}
\end{figure}

\begin{figure}
  \centering
  \includegraphics[width=0.5\textwidth]{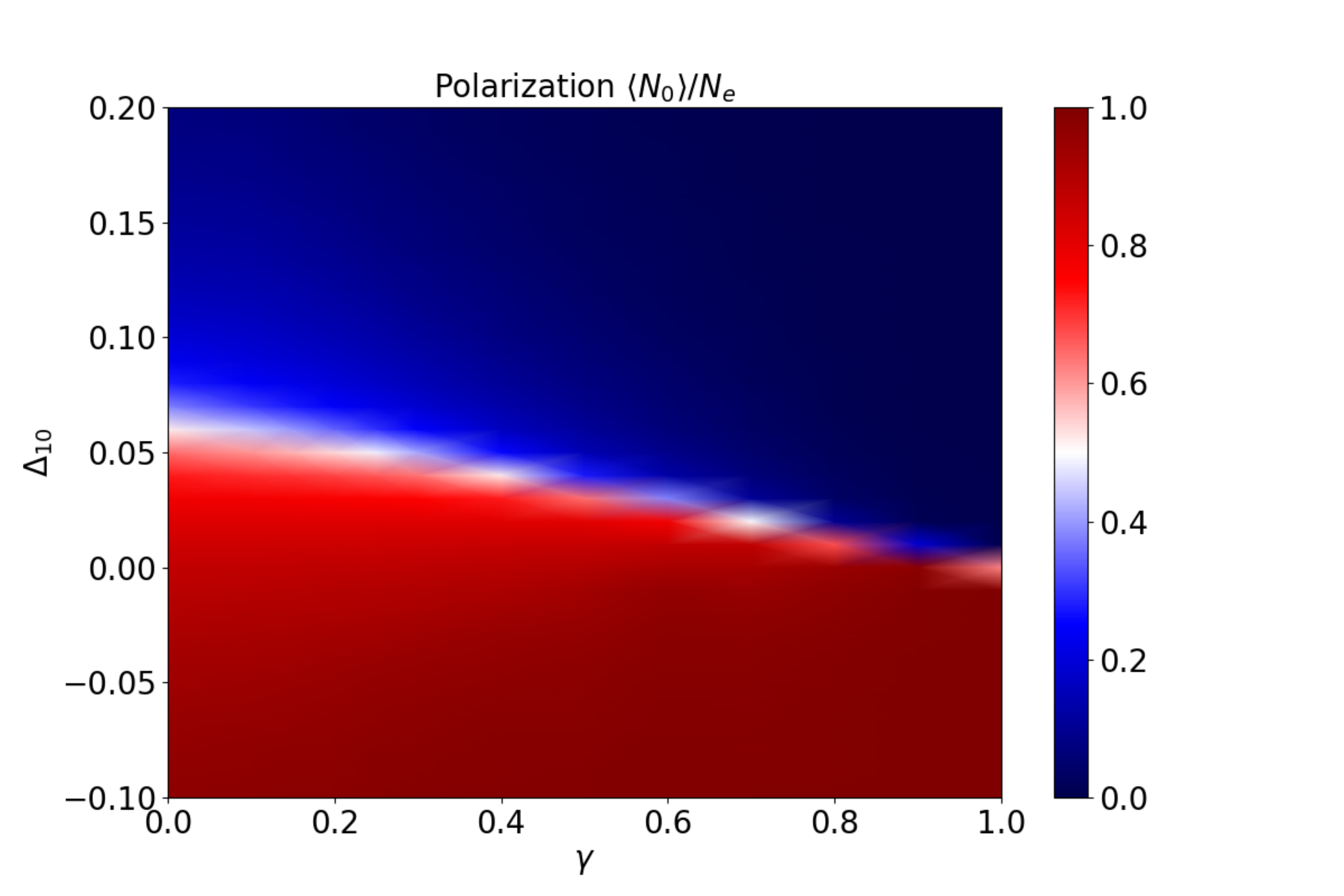}
  \caption{Polarization of the zero-momentum ground state for N=6 electrons as a function of the $\gamma$ parameter
  and the splitting $\Delta_{10}$ between the two levels. In the $SU(2)$ symmetric case for $\gamma=1$ the
  transition happens right at $\Delta_{10}=0$ while it is shifted in the upper half plane when $\gamma <1$}
  \label{polarmap}
\end{figure}

It is natural to define a polarization of the two-component system 
of any quantum state $\Psi$ by the following quantity~:
\be
\mathcal{P}=\langle\Psi|{\hat N}_0|\Psi\rangle/N_e,
\ee
that will track the orbital content of the state under consideration.
With the normalization we have chosen $\mathcal{P}=1$ corresponds to all electrons
residing in the $N=0$ orbital.
We infer that for any state $\mathcal{P}$ will vary from +1 in the lower part of the phase diagram 
for $\Delta_{10}$ negative to zero in the upper part for $\Delta_{10}$ positive.
Exactly on the $\gamma=1$ line the behavior of $\mathcal{P}$ can be inferred from the nature
of the spin multiplet of the ferromagnetic Laughlin state. The curve is a simple staircase with two values
according to the discussion about the quantum Hall ferromagnetism above. For $\gamma <1$ the curve is 
non-trivial.
We find that the polarization curve is now rounded and has a sharp decrease when $\Delta_{10}$ goes up
in all parts of the phase diagram. However the inflection point of the curve is no longer pinned at 
$\Delta_{10}=0$ but is at positive nonzero values for small enough $\gamma$. We display in Fig.(\ref{polarplot})
the case of $\gamma=0$ which is the leftmost boundary of the phase diagram.
We find that the sharp transition happens for $\Delta_{10}\approx 0.06$ almost independently of the size of the
system. When $\gamma$ is nonzero the sharp transition happens for decreasing values of $\Delta_{10}$.
A map of the polarization is given in Fig.(\ref{polarmap}). This is expected since the exchange energy
is stronger in the lowest Landau level with respect to the second Landau level. So even when $\Delta_{10}=0$
it is more favorable energetically to polarize the system in the lowest Landau level. One needs a finite
amount of offset energy $\Delta_{10}$ to overcome the exchange energy gain.

In the complete phase diagram we find a phase boundary given in Fig.(\ref{polarmap}) where we see the
polarization boundary evolving smoothly.

\subsection{Using fidelity to locate phase boundaries}

\begin{figure}[h]
\centering
 \includegraphics[width=0.5\columnwidth]{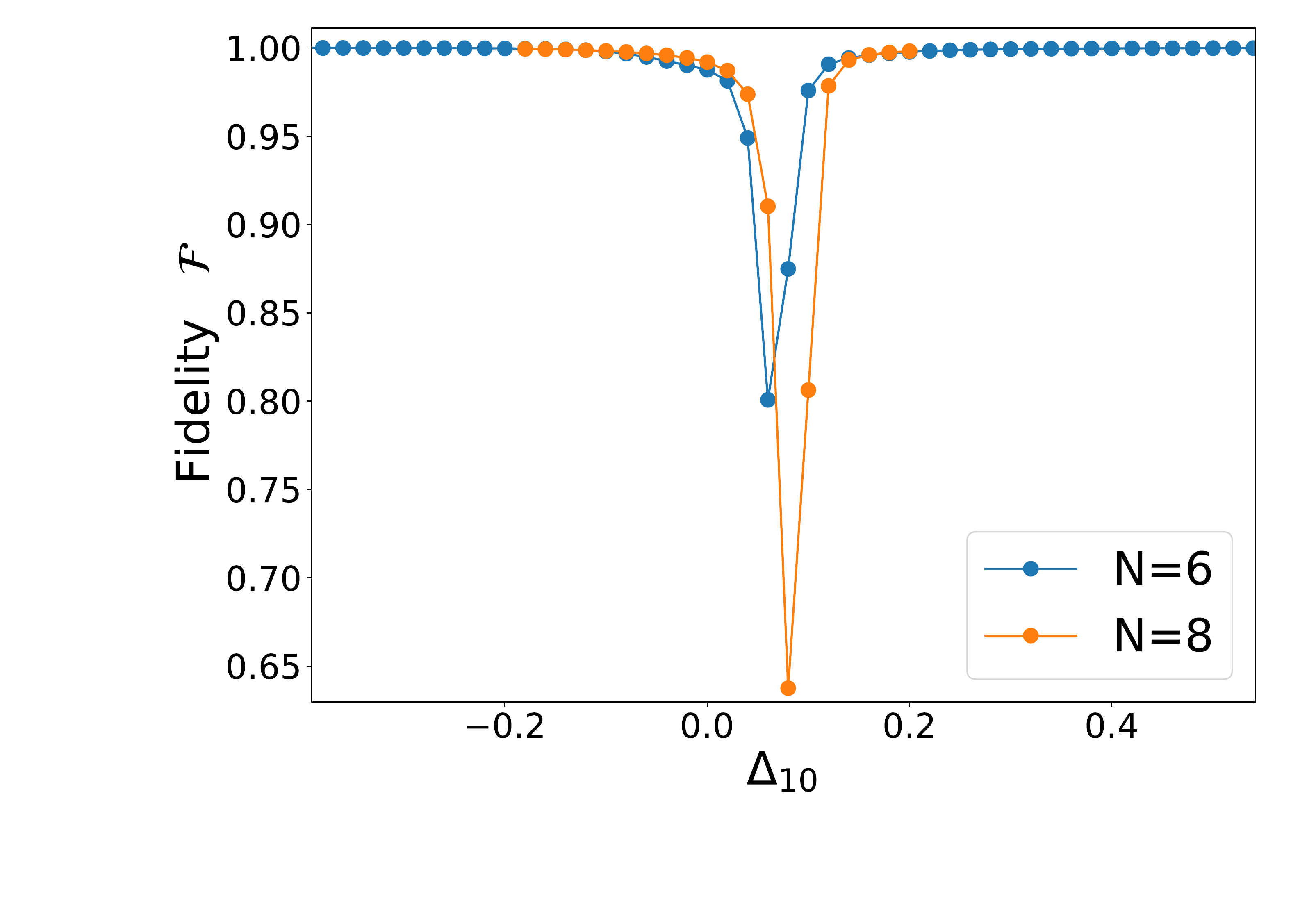}
 \caption{The fidelity 
 at $\nu=1/3$ for  $\gamma=0$ as a function of $\Delta_{10}$. We track the zero-momentum ground state.
 There is evidence for a transition between two regimes happens 
 for a non-trivial value of the bare splitting $\Delta_{10}\approx 0.06$ weakly sensitive to the system sizes 
 we have studied. This is numerically equal or very close to the value of maximal slope of the
 polarization curve }
 \label{fidelplot}
\end{figure}

To locate phase boundaries one can also compute overlaps of different ground states
obtained by changing a parameter of the Hamiltonian. An infinitesimal version of the overlap calculation
is the so-called quantum fidelity~:
\be
\mathcal{F}(\delta)=|\langle \Psi(\delta+\epsilon)|\Psi(\delta)\rangle|,
\label{fidelitydef}
\ee
where $\delta$ is the parameter we choose to vary and $\epsilon$ an infinitesimal increment.
Phase transitions are expected to appear as a strong dip away from unity and the bigger the size
the stronger the dip. We have computed the fidelity across the phase diagram and it is sensitive
to the polarization transition what we observed above. A sample behavior at fixed gamma as a function
of $\Delta_{10}$ is given in Fig.(\ref{fidelplot}). The characteristic dip grows with the size of the system
as expected for a phase transition.
With only these two indicators there is no sign of other phase boundaries.

\subsection{The Wigner crystal}

\begin{figure}[h]
\begin{center}
  \vcenteredhbox{\includegraphics[width=0.45\textwidth]{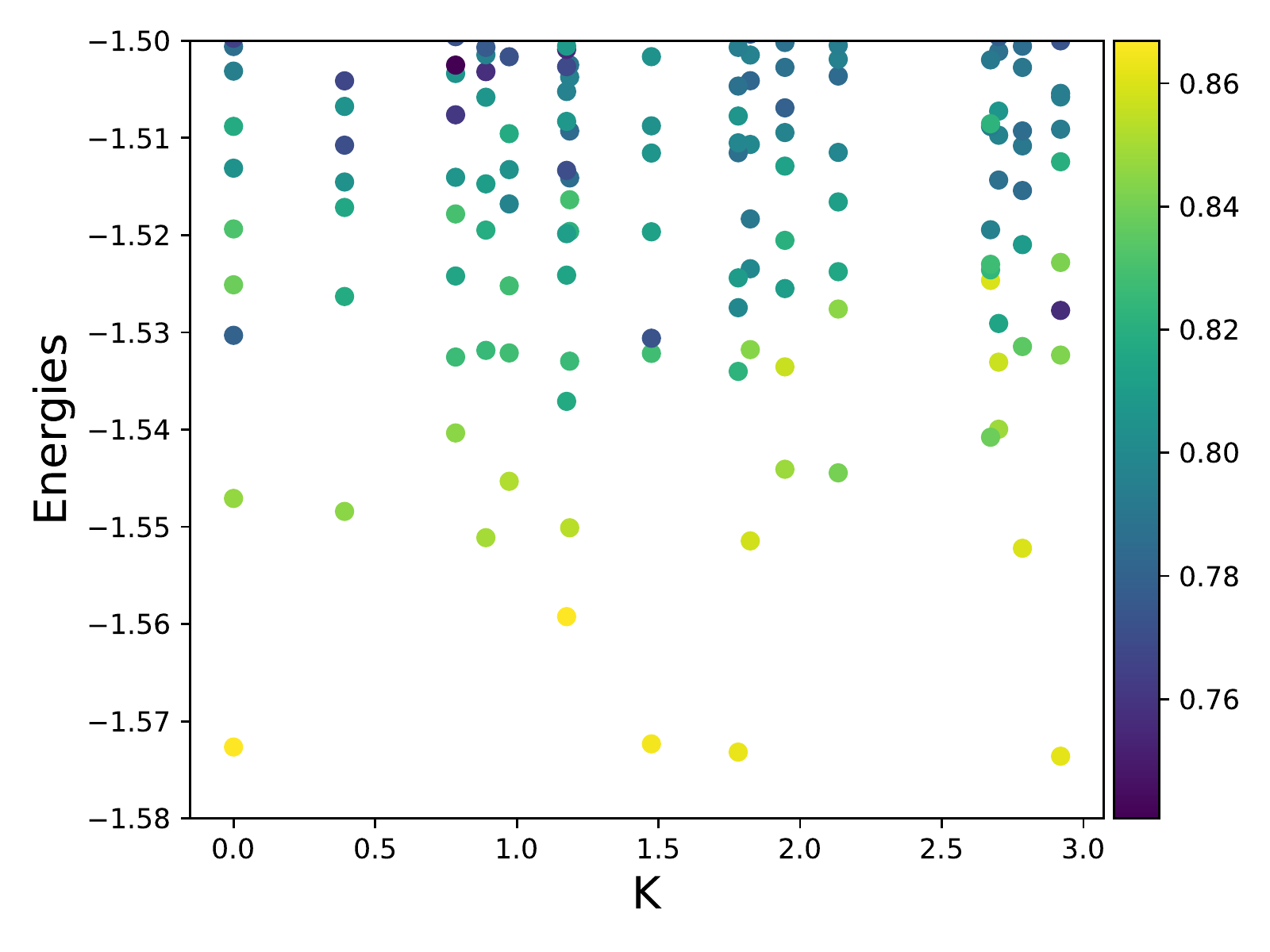}}
  \hspace{1em}
  \vcenteredhbox{\includegraphics[width=0.45\textwidth]{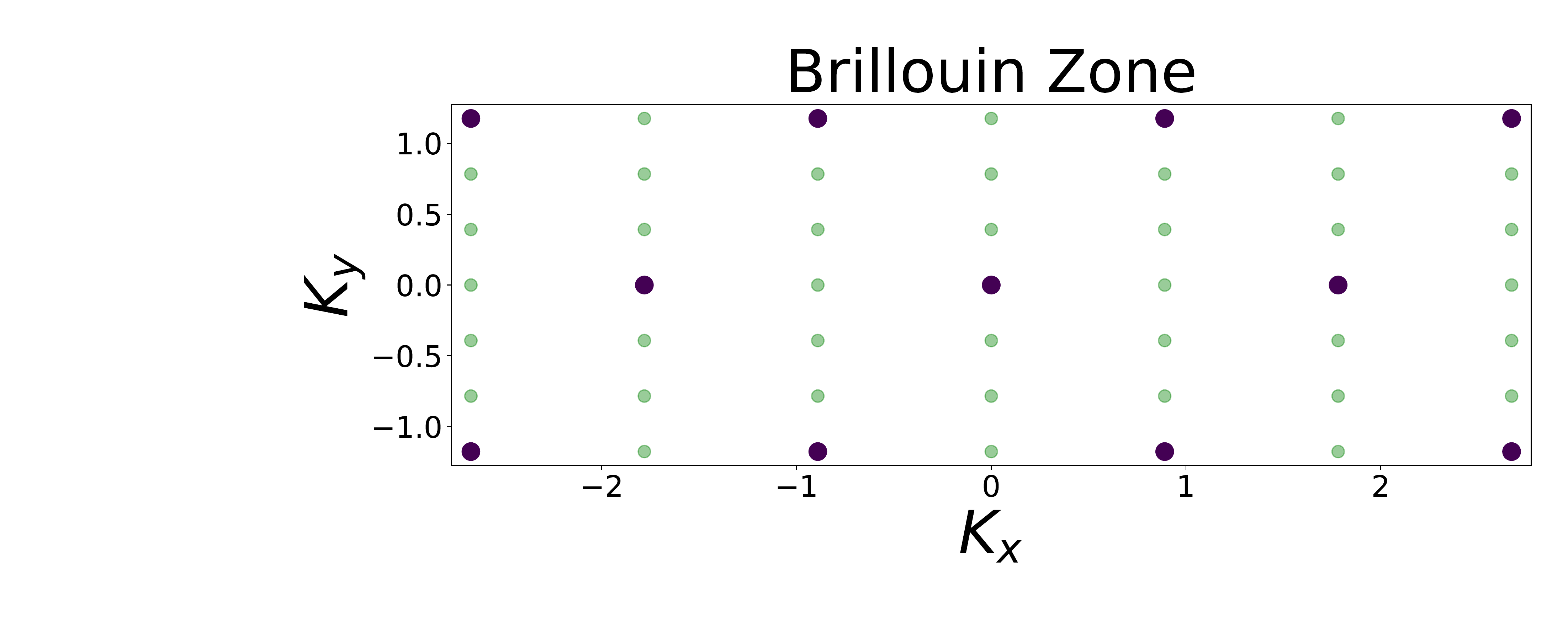}}
\end{center}
 \caption{Energy spectrum for 6 electrons with $\gamma=0$ and exact degeneracy $\Delta_{10}=0$.
 The energies are plotted as a function of the momentum $K$ in a rectangular cell
 of aspect ratio 0.44. The color indicates the polarization of the low-lying eigenstates.
 There is a set of quasidegenerate states clearly separated from the reminder of the spectrum.
 They are all essentially fully polarized in the N=0 orbital subspace.}
 \label{GSmanifold}
 \end{figure}

Beyond the simple two-phase picture given by the polarization and fidelity transition of the section above
we nevertheless find that the physics changes drastically in the region of small $\gamma$ and small 
$\Delta_{10}$. Indeed we observe the appearance in the spectrum of a set
of nearly-degenerate low-energy states well separated from higher-energy states
by a gap. We call these states the ground state manifold. 
 One has to fine tune the aspect ratio
of the periodic unit cell used in the diagonalization~\cite{HRY1999,HRY2000,YHR2001}
to obtain the degeneracy.
This is a typical occurrence of broken symmetry
in a finite system. This phenomenon has been observed in the lowest Galilean Landau level
for small filling factors and is an evidence of a Wigner crystal ground state as found in 
refs.(\onlinecite{HRY1999,HRY2000,YHR2001}). An example is displayed in the right panel of
Fig.(\ref{GSmanifold})
where the energy eigenvalues are plotted as a function of the momentum. The set of almost degenerate
states has an energy splitting of the order of at most $\approx 10^{-4}$ in Coulomb scale units
as opposed to excitation energies of $O( 10^{-2})$. If plotted in the Brillouin zone the momenta
of the quasidegenerate states display a regular lattice which is the reciprocal lattice of a
crystal structure as seen in the right panel of Fig.(\ref{GSmanifold}). In most cases we have studied the 
lattice is almost triangular.
Also the number of states in the quasidegenerate manifold is equal to the number of electrons,
ruling out bubble or charge density wave states~\cite{HRY1999,HRY2000,YHR2001}.

 \begin{figure}[h]
\centering
 \includegraphics[width=0.5\columnwidth]{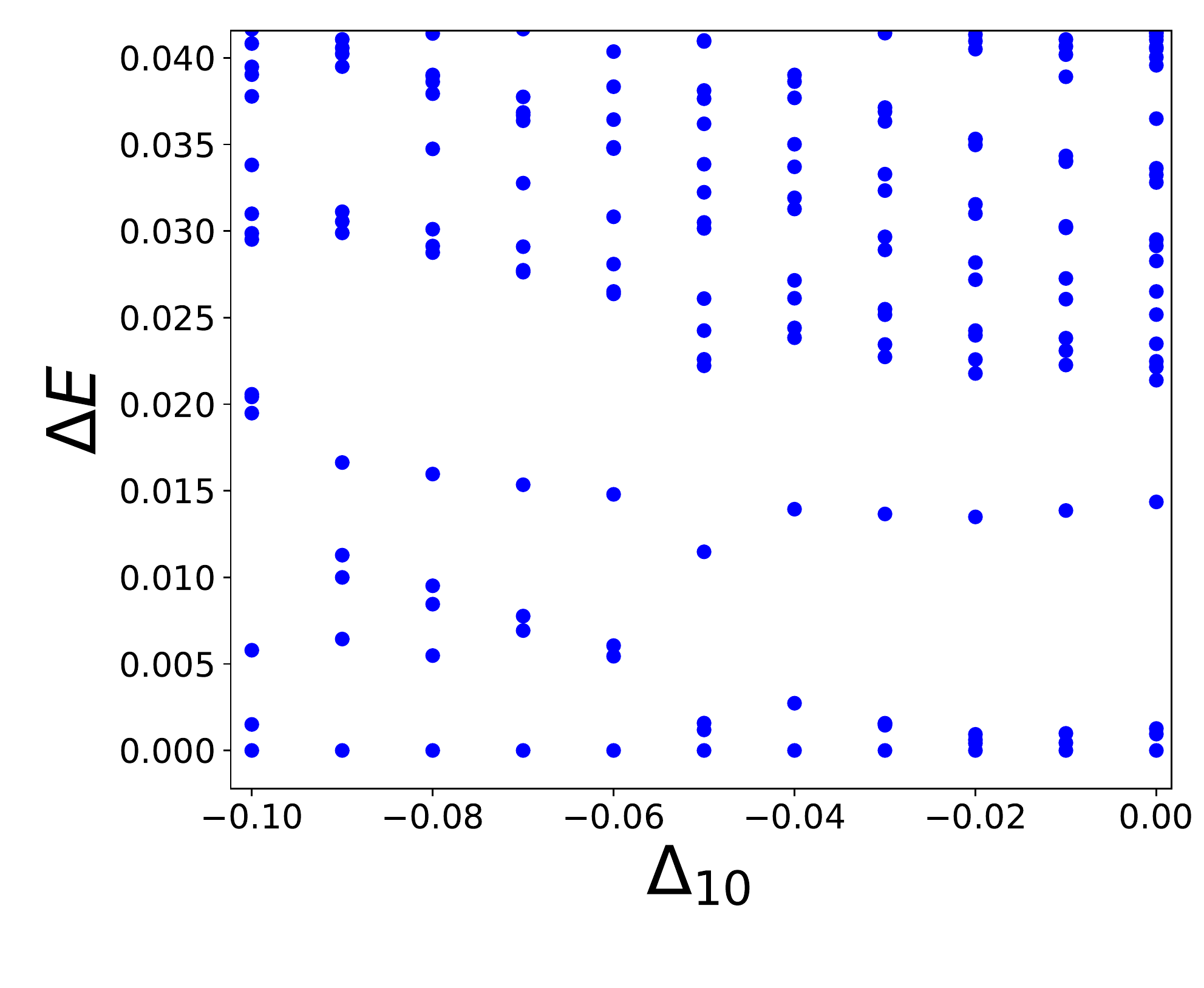}
 \caption{Excitation energies above the ground state as a function of $\Delta_{10}$
 for $\gamma=0$ for a system of $N_e=6$ particles at $\nu=1/3$ in a rectangular unit cell of aspect ratio
 $L_x/L_y=0.44$. When there is orbital degeneracy $\Delta_{10}=0$ we observe a set of quasidegenerate states
 pointing to broken translational symmetry. An increase of the pseudo-Zeeman field lifts the degeneracies in a smooth crossover starting around $\Delta_{10}\approx -0.02$}
 \label{degeneracies}
\end{figure}

This manifold of states is sensitive to the aspect ratio $L_x/L_y$
of the unit cell in which we diagonalize the Hamiltonian. There is a optimal value which is size dependent
as is the case of the low-filling factor Wigner crystal~: $L_x/L_y=0.44$ for $N_e=6$, $0.37$ for $N_e=7$,
$0.7$ for $N_e=8$, $0.3$ for $N_e=9$. 
From observation of spectra like the ones presented
in Fig.(\ref{degeneracies}) it is difficult to pinpoint a phase boundary separating the Wigner crystal from
the Laughlin state. We just observe a crossover with no striking discontinuity or (obvious) non-analyticity.
An example of the crossover is given in Fig.(\ref{degeneracies}).
For positive splitting $\Delta_{10}$ the crystal does not survive the polarization transition
and the degeneracy is destroyed. If we increase $\gamma$ we also find that the crystal signature disappears
close to $\gamma\approx 0.7$. 

\begin{figure}[h]
\centering
 \includegraphics[width=0.3\columnwidth]{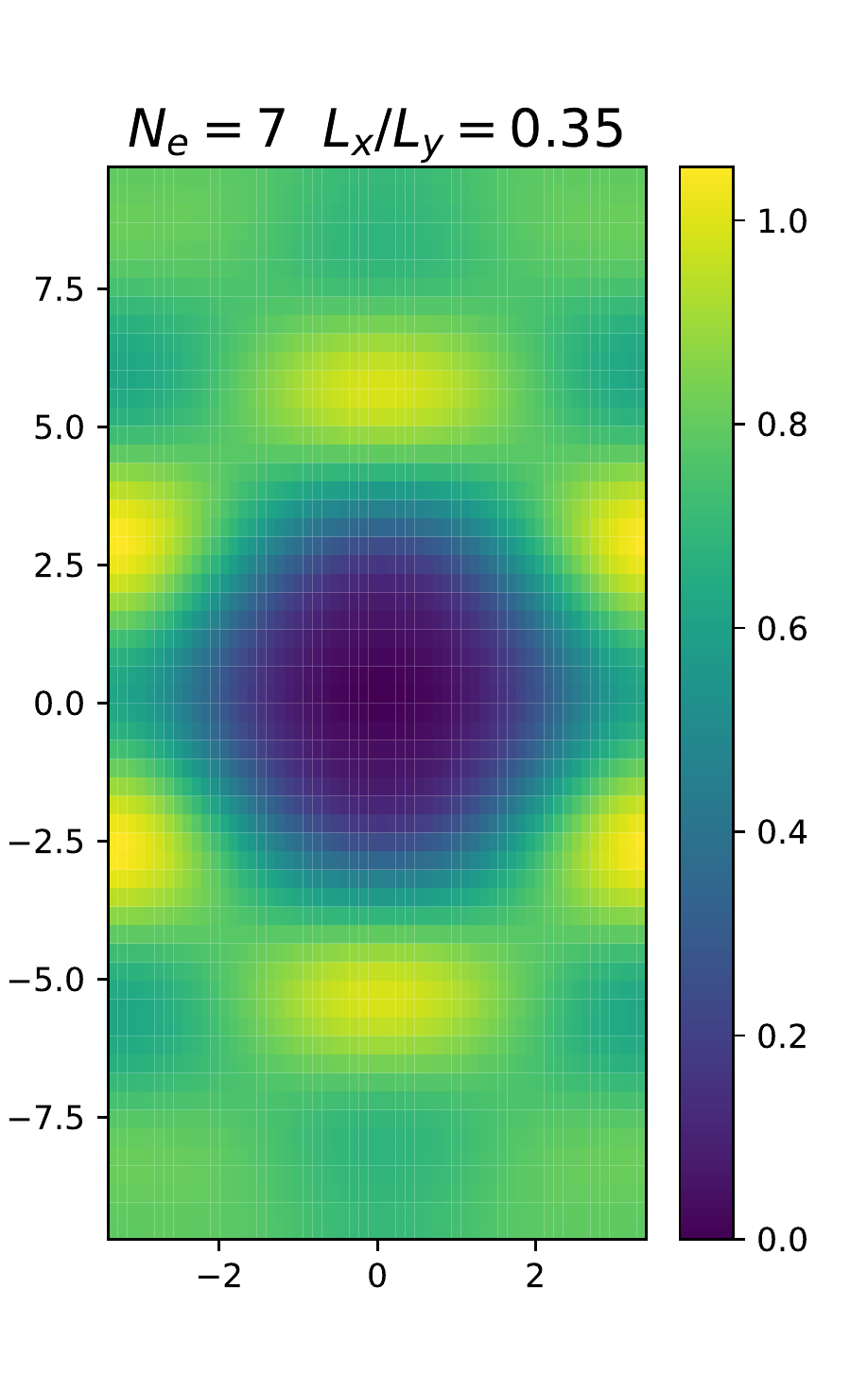}
\includegraphics[width=0.3\columnwidth]{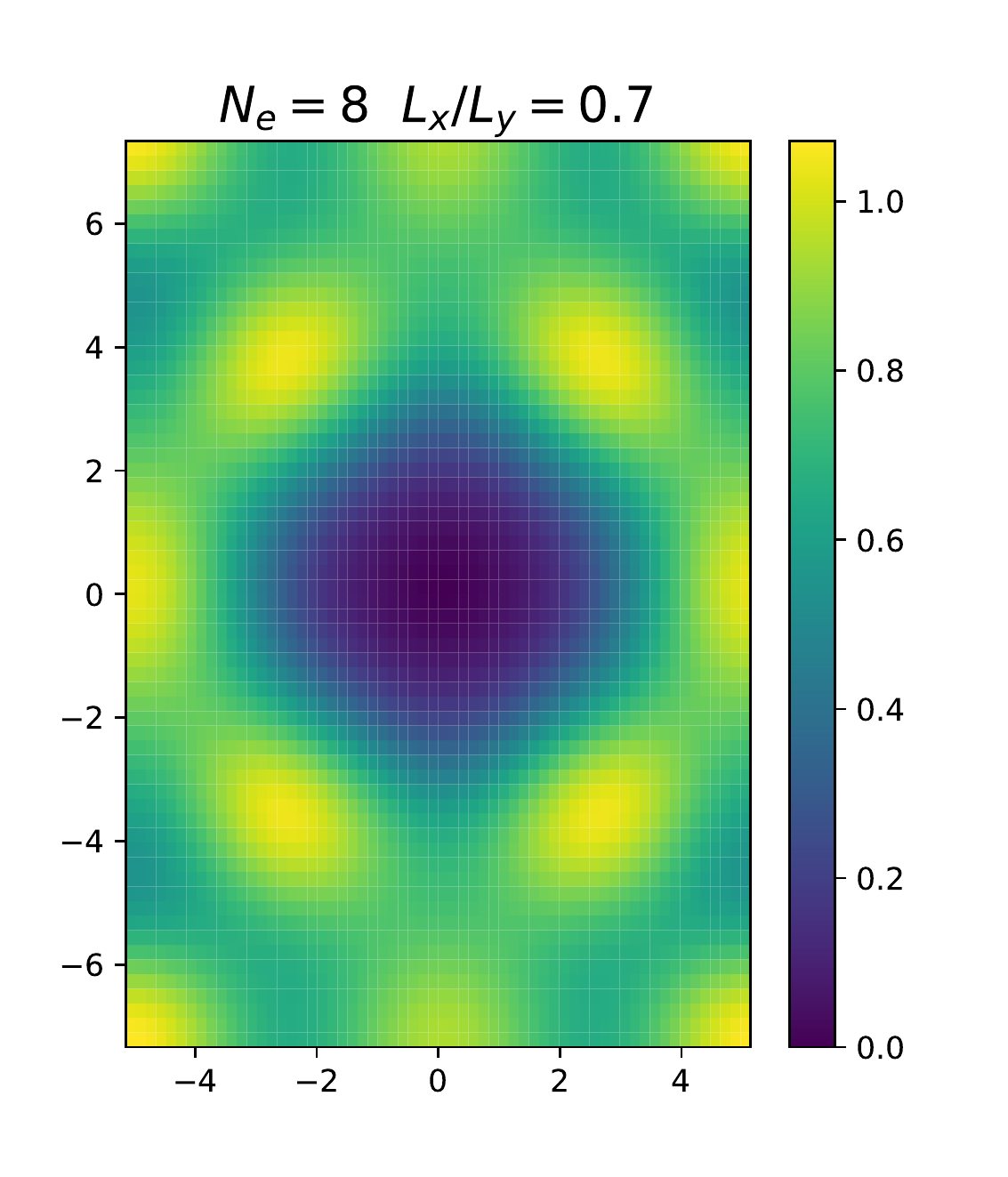}
\includegraphics[width=0.3\columnwidth]{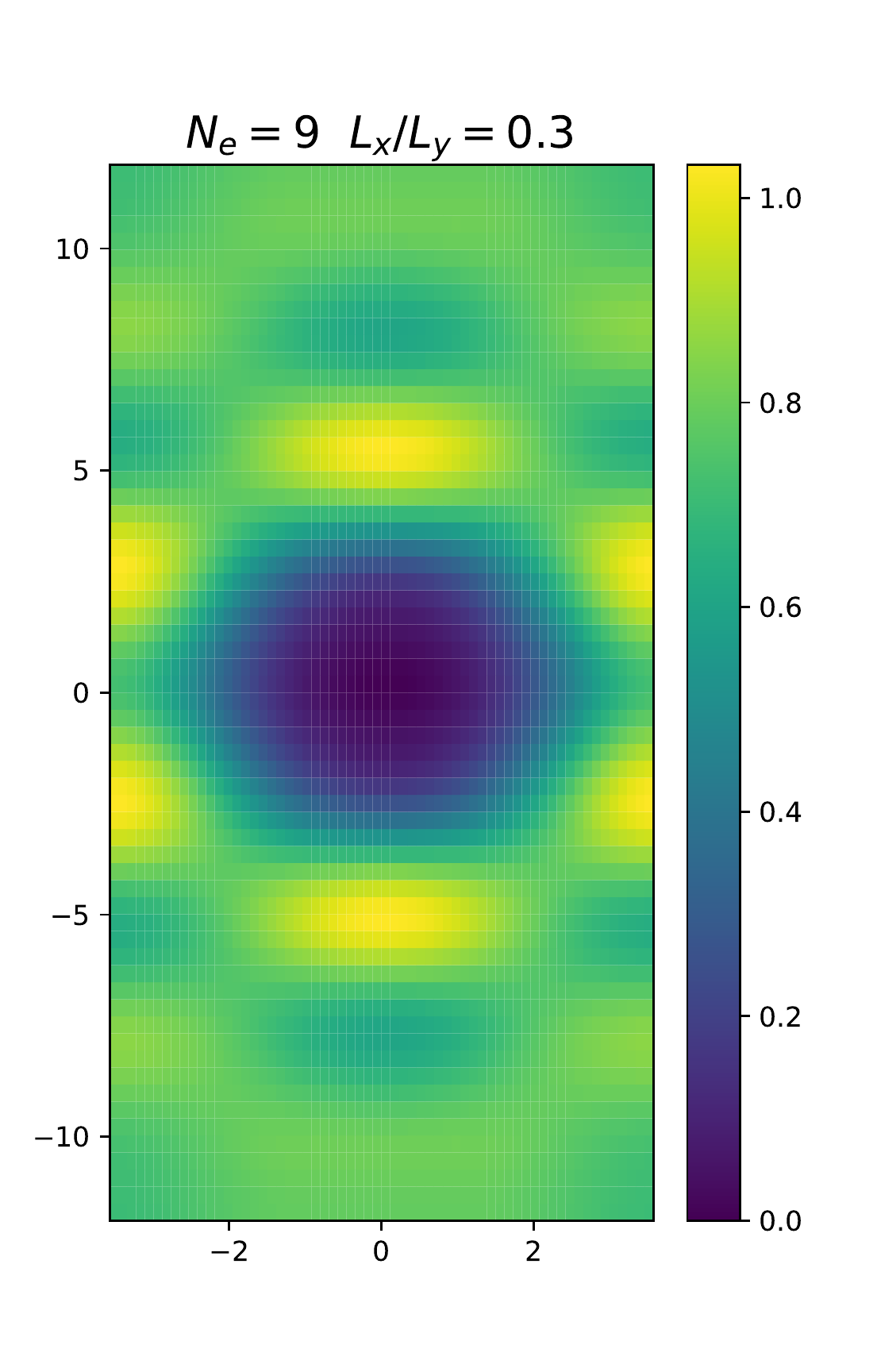}
 \caption{The pair correlation function $g_{00}(x,y)$ for $N_e=7$ (left),8 (center),9 (right) electrons
 at $\nu=1/3$ for $\Delta_{10}=0$ and $\gamma=0$. The electron located at the origin
 gives a deep correlation hole in addition to the Pauli zero. The aspect ratio is chosen so
as  to favor the appearance of the crystal structure. This value of $L_x/L_y$ is size dependent.
For $N_e=7,9$ (left and right) there is evidence for a triangular crystal state while it is a square lattice
that appears for $N_e=8$ (center). We note that revealing the square case requires a different aspect ratio
than for the triangular case.}
 \label{pair-xtal}
\end{figure}

To characterize the crystalline order we also compute the pair correlation function~:
\be
g_{\alpha\beta}(R)=\frac{1}{\rho N_e}\sum_{i\neq j}\delta(r_i-r_j-R)(|\alpha\rangle\langle\alpha|)_i
(|\beta\rangle\langle\beta|)_j
\label{paircorr}
\ee
where $\alpha,\beta=0,1$ is the orbital index and electron coordinates are $r_i,r_j$, $\rho$ is the density.
In this definition the pair correlation is periodic 
$g_{\alpha\beta}(x+L_x,y)=g_{\alpha\beta}(x,y)$, 
$g_{\alpha\beta}(x,y+L_y)=g_{\alpha\beta}(x,y)$.
The almost full polarization of the crystal states means that only $g_{00}$ is sizeable.
Indeed we find that other components $g_{11}$,$g_{10}$,$g_{01}$ track the pattern seen in $g_{00}$.
In the region of degeneracies we observe directly a crystal structure
in real space from $g_{00}$ as displayed in Fig.(\ref{pair-xtal}). A liquid state
has instead a single crater-like feature around the position of the reference particle
and a smooth background beyond a first ring of overdensity. Here what we observe is definitely
a state modulated in real space and the lattice of electrons has the reciprocal lattice
we observe in the manifold of degenerate states. We take this as evidence for a crystal state.
The number of overdensities plus the central electron is equal to the total number of electrons,
again consistent with a Wigner crystal. The case of $N_e=8$ is special~: the lattice formed
is square as can be seen in the central panel of Fig.(\ref{pair-xtal}) and the momenta in the Brillouin zone
also form a square lattice. While we cannot ultimately conclude on the precise nature of the lattice,
this is indicative of a close competition between triangular and square lattices.

\begin{figure}[h]
\centering
\includegraphics[width=0.3\columnwidth]{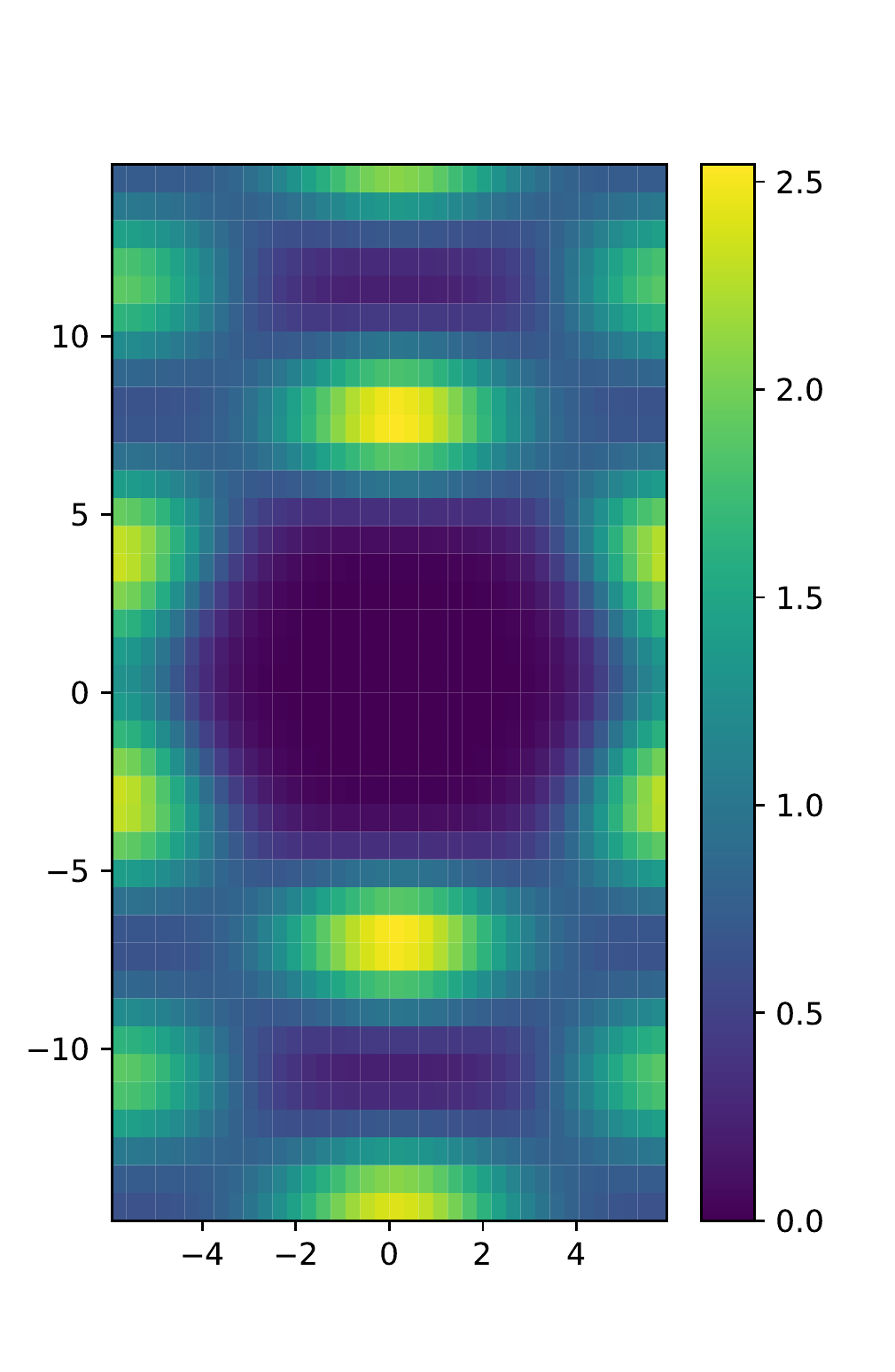}
 \caption{The pair correlation function $g(x,y)$ for $N=8$  electrons
 in the lowest Galilean Landau level
 at filling factor $\nu=1/7$. The aspect ratio is taken to
 be $L_x/L_y=0.4$ to favor the appearance of the crystal structure as was observed for smaller system sizes
 in refs.(\onlinecite{HRY1999,HRY2000,YHR2001}). There are eight overdensities which is consistent with a Wigner
 crystal with one electron per site and a triangular structure. The density contrast is much higher for this
 system size and aspect ratio than in the case of the bilayer system.}
 \label{pair-xtal-smallfilling}
\end{figure}

To characterize the crystal strength we can measure the contrast around the first ring of overdensity
in the pair correlation function
which is prominent both in solid and liquid phases. In the crystal state we find a contrast
$(g_{max}-g_{min})/g_{min}\approx 0.44$ when $\gamma=0$ and $\Delta_{10}=0$ for $N_e=9$. If we consider the
 liquid Laughlin state the contrast is only $0.07$ for a system of the same size. The Coulomb ground state
 at $\nu=1/3$ in a single polarized Landau level, well described by the Laughlin state, has even less
 contrast $\approx 0.04$. We use this quantity to pin down the boundary of the crystal phase in the region
 with negative $\Delta_{10}$. We decide quite arbitrarily that the state is liquid when the contrast
 is less than $0.1$. This gives a rough estimate of the stability region, leading to the red island 
 in Fig.(\ref{PhaseDiag}). As a comparison we display in Fig.(\ref{pair-xtal-smallfilling}) the pair
 correlation for one-component spin-polarized electrons in the lowest Galilean Landau level for $N_e=8$
 electrons at filling factor $\nu=1/7$ where we expect a Wigner crystal~\cite{HRY1999}. Indeed 
 the contrast is now close 
 to $2$ along the ring circling the reference electron at $r=0$. We attribute this difference to the fact 
 that the unit cell is now larger in real space for the fraction $\nu=1/7$~: this case has $N_\phi=56$ 
 leading to sizes $L_x=11.9$ 
 and $L_y=29.6$ while our largest system for $N_e=9$ has $L_x=7.1$ and $L_y=23.8$ so finite size effects
 are presumably stronger for the bilayer case.

Finally we note that the Wigner state appears to be almost fully polarized~: see the color bar in 
Fig.(\ref{GSmanifold}) as a representative example. An interesting question is whether this state is 
totally polarized $\mathcal{P}=1$ or almost fully polarized $\mathcal{P}\lesssim 1$. From our numerical
measurements we cannot discriminate between these two possibilities. With the accessible sizes
finite size scaling is not conclusive.
Partial polarization would be
an interesting new phenomenon. Non-trivial modifications of existing trial wavefunctions would be required
to explain such a state.

\section{Behavior of the $\nu=2/3$ state}
\label{TwoThirds-section}
We now turn to the study of the FQHE state at filling factor $\nu=2/3$. Due to the two-component
nature of the BLG states this is not simply the particle-hole conjugate of the $\nu=1/3$ case
which happens indeed at $\nu=2-1/3=5/3$.
We start by discussing the physics for $\gamma=1$ as a function of the external field 
$\Delta_{10}$. In this limiting case we have full $SU(2)$ symmetry in the orbital space
only broken by the Zeeman-like field $\Delta_{10}$.
For two-component FQHE states at $\nu=2/3$ we know that there are two incompressible candidate
states that compete. The first candidate is the particle-hole conjugate of the $\nu=1/3$ state
where the particle-hole symmetry does not involve the spin degrees of freedom. 
It can be viewed as a composite fermion state with negative effective flux
and two filled effective Landau levels~\cite{JainBook}.
This
state is fully polarized and in the absence of Zeeman energy it gives rise to an exactly
degenerate 
ferromagnetic spin multiplet.
The other candidate state is the $\nu=2/3$ global spin singlet state which is known to be
the ground state in the absence of Zeeman energy i.e. it is definitely lower than the fully polarized
2/3 state. For $\gamma=1$ there is a sharp transition between these two ground states as a
function of $\Delta_{10}$ by translation of the known spin physics.

If we now  consider the realistic case of $\gamma <1$, the $SU(2)$ symmetry is explicitly broken
and nothing precludes a continuous transformation between these two states.
This is indeed what we observe in our diagonalizations. We find no transitions in the interior region
of the phase diagram. At $\gamma=1$ and varying $\Delta_{10}$ the polarization $\mathcal{P}$
has two sharp transitions from +1 to 1/2 and then to 0 which correspond to the expected spin transitions
of $\nu=2/3$, the plateau at $\mathcal{P}=1/2$ corresponding to the singlet state. For $\gamma <1$ 
these steps are rounded and no transitions remain. We conclude that there is no Wigner 
crystal state at $\nu=2/3$.

\section{Conclusions}
\label{concluding}

We have studied by exact diagonalization techniques in the torus geometry
the fate of the quantum Hall states at $\nu=1/3$ and $\nu=2/3$ in the bilayer graphene
system when there is almost coincidence of Landau levels with N=0 and N=1 orbital character.
The detailed quantitative studies of ref.(\onlinecite{Hunt}) have shown that this level coincidence can happen for $\nu_{BLG}=-3+1/3,-3+2/3$ notably. Previous studies of this coincidence~\cite{Sodemann}
have shown a complex competition of phases for the half-filled case.
By tuning the applied magnetic field and the electric bias between the layers we find that
it is possible to destabilize the Laughlin incompressible state for filling 
$\nu=1/3$ and create a Wigner crystal of electrons.
This crystal is stabilized when the N=0 and N=1 are in almost coincidence
creating a situation akin to extreme Landau level mixing~\cite{Zhao} with the difference
that there are no extra levels beyond N=1. The crystal structure is revealed by the quasidegeneracies
in the many-body spectrum as well as by the pair correlation function. The crystal is favored by fine-tuning
the rectangular unit cell we use in exact diagonalizations. The crystal structure seen in real space
from the measurement of the pair correlation has the same reciprocal lattice as observed in the magnetic Brillouin zone for the quasidegenerate states. There is a polarisation transition that coincides
with the boundary of the crystal phase for positive values of $\Delta_{10}$ at least when $\gamma$ is small enough. For negative values of the level splitting we use the contrast of the ring of overdensity
of the pair correlation to locate the other phase boundary. The resulting map of the crystal state is given
in our Fig.(\ref{PhaseDiag}). The simplest way to observe the transition is to measure the longitudinal
resistance that should have a sharp peak when there is level coincidence. 
Thermodynamic measurements like the chemical potential can be used~\cite{Yang-MLG11}.
The interlayer bias should be large enough while there is no real restriction of the value of the magnetic field which controls the $\gamma$ parameter. The phase transition betwen the crystal state and the Laughlin state involves only a smooth
lifting of the degeneracy of the ground state multiplet and the zero-momentum state is always a member
of the degenerate states, deforming continuously into the Laughlin state. This is compatible with
a weakly first order or second order transition.

\begin{acknowledgments}
We acknowledge discussions with A. Assouline and thank M. Shayegan for useful correspondence.
We thank DRF and GENCI-CCRT for 
computer time allocation on the Topaze cluster.
\end{acknowledgments}



\begin{thebibliography}{99}


\bibitem{Du} 
X. Du, I. Skachko, F. Duerr, A. Luican, and E. Y. Andrei,
Nature 462, 192 (2009).

\bibitem{Bolotin} 
K. I. Bolotin, F. Ghahari, M. D. Shulman, and H. L. Stormer, P. Kim, 
Nature 462, 196 (2009).

\bibitem{Ghahari} 
F. Ghahari, Y. Zhao, P. Cadden-Zimansky, and K. Bolotin, P. Kim, 
Phys. Rev. Lett. {\bf 106}, 046801 (2011).

\bibitem{Dean-MLG1} 
C. R. Dean, A. F. Young, P. Cadden-Zimansky, L. Wang, H. Ren, K. Watanabe, T. Taniguchi, 
P. Kim, J. Hone, and K. L. Shepard, 
Nature Physics {\bf 7}, 693 (2011).

\bibitem{Young-MLG2} 
A. F. Young, C. R. Dean, L. Wang, H. Ren, P. Cadden-Zimansky, K. Watanabe, 
T. Taniguchi, J. Hone, K. L. Shepard, and P. Kim, 
Nature Phys. {\bf 8}, 550 (2012).

\bibitem{Feldman-MLG3} 
B. E. Feldman, B. Krauss, J. H. Smet, and A. Yacoby, 
Science {\bf 337}, 1196 (2012).

\bibitem{Feldman-MLG4} 
B. E. Feldman, A. J. Levin, B. Krauss, D. A. Abanin, B. I. Halperin, J. H. Smet, and 
A. Yacoby, 
Phys. Rev. Lett. {\bf 111}, 076802 (2013).

\bibitem{Hunt}
B. Hunt, J. D. Sanchez-Yamagishi, A. F. Young, M. Yankowitz, B. J. LeRoy, K. Watanabe,
T. Taniguchi, P. Moon, M. Koshino, P. Jarillo-Herrero, and R. C. Ashoori, 
Science {\bf 340}, 1427 (2013).

\bibitem{Young-MLG4} 
A. F. Young, J. D. Sanchez-Yamagishi, B. Hunt, S. H. Choi, K. Watanabe, T. Taniguchi, 
R. C. Ashoori, and P. Jarillo-Herrero, 
Nature {\bf 505}, 528 (2014).

\bibitem{Amet-MLG5}
F. Amet, A. J. Bestwick, J. R. Williams, L. Balicas, K. Watanabe, T. Taniguchi,
and D. Goldhaber-Gordon,
Nature Commun. 6:5838 (2015).

\bibitem{Zibrov-MLG6}
A. A. Zibrov, E. M. Spanton, H. Zhou, C. Kometter, T. Taniguchi, K. Watanabe, 
and A. F. Young,
Nature Physics, \textbf{14}, 930 (2018).

\bibitem{Polshyn-MLG7}
H. Polshyn, H. Zhou, E. M. Spanton, T. Taniguchi, K. Watanabe, 
and A. F. Young,
Phys. Rev. Lett. {\bf 121}, 226801 (2018).

\bibitem{Chen-MLG8}
S. Chen, R. Ribeiro-Palau, K. Yang, K. Watanabe, T. Taniguchi,
J. Hone, M. O. Goerbig, and C. R. Dean,
Phys. Rev. Lett. {\bf 122}, 026802 (2019).

\bibitem{Zeng-MLG9}
Y. Zeng, J. I. A. Li, S. A. Dietrich, O. M. Ghosh, K. Watanabe, T. Taniguchi,
J. Hone, and C. R. Dean,
Phys. Rev. Lett. {\bf 122}, 137701 (2019).

\bibitem{Zhou-MLG10}
H. Zhou, H. Polshyn, T. Taniguchi, K. Watanabe, 
and A. F. Young,
Nature Physics \textbf{16}, 154 (2020).

\bibitem{Yang-MLG11}
F. Yang, A. A. Zibrov, R. Bai, T. Taniguchi, K. Watanabe, M. P. Zaletel, 
and A. F. Young,
Phys. Rev. Lett. \textbf{126}, 156802 (2021).

\bibitem{Apalkov2010}
V. M. Apalkov and T. Chakraborty,
Phys. Re. Lett. \textbf{105}, 036801 (2010).

\bibitem{Zibrov2017}
A. A. Zibrov, C. Kometter, H. Zhou, E. M. Spanton, T. Taniguchi , K. Watanabe, M. P. Zaletel, and A. F. Young,
Nature \textbf{549}, 360 (2017).

\bibitem{Maher-BLG1}
P. Maher, C. R. Dean, A. F. Young, T. Taniguchi, K. Watanabe, K. L. Shepard, J. Hone, and P. Kim,
Nature Physics \textbf{9}, 154 (2013).

\bibitem{Maher-BLG2}
P. Maher, L. Wang, Y. Gao, C. Forsythe, T. Taniguchi, K. Watanabe, D. Abanin, Z. Papic, 
P. Cadden-Zimansky, J. Hone, P. Kim, C. R. Dean
Science \textbf{345}, 61 (2014).

\bibitem{Kou-BLG3}
A. Kou, B. E. Feldman, A. J. Levin, B. I. Halperin,  K. Watanabe, T. Taniguchi,
A. Yacoby,
Science \textbf{345}, 55 (2014).

\bibitem{Li-BLG4}
J. I. A. Li, C. Tan, S. Chen, Y. Zeng, T. Taniguchi, K. Watanabe, J. Hone, C. R. Dean,
Science \textbf{358}, 648 (2017).

\bibitem{Pientka-BLG5}
F. Pientka, J. Waissman, P. Kim, and B. I. Halperin,
\phrl{119}{027601}{2017}.

\bibitem{Hunt-BLG6}
B. M. Hunt, J. I. A. Li, A. A. Zibrov, L. Wang, T. Taniguchi, K. Watanabe, J. Hone, C. R. Dean,
M. Zaletel, R. C. Ashoori, and A. F. Young,
Nature Communications \textbf{8:948} (2017).

\bibitem{Fu-BLG7}
H. Fu, K. Huang, K. Watanabe, T. Taniguchi, and J. Zhu,
Phys. Rev. X\textbf{11}, 021012 (2021).

\bibitem{Seiler-BLG8}
A. M. Seiler, F. R. Geisenhof, F. Winterer, K. Watanabe, T. Taniguchi, T. Xu, F. Zhang, R. T. Wietz,
Nature 608 (2022) 298.

\bibitem{Huang2022}
Ke Huang, Hailong Fu, Danielle Reifsnyder Hickey, Nasim Alem, Xi Lin, Kenji Watanabe, 
Takashi Taniguchi, and Jun Zhu,
Phys. Rev. X{\bf 12}, 031019 (2022).

\bibitem{Laughlin}
R. B. Laughlin,
\phrl{50}{1395}{1983}

\bibitem{LeGlatt}
E. Y. Andrei, G. Deville, D. C. Glattli, F. I. B. Williams, E. Paris, and B. Etienne,
Phys. Rev. Lett. \textbf{60}, 2765 (1988).

\bibitem{Zuo2020}
Z.-W. Zuo, A. C. Balram, S. Pu, J. Zhao, Th. Jolicoeur, A. Wójs, and J. K. Jain,
Phys. Rev. B\textbf{102}, 075307 (2020).

\bibitem{Shayegan98}
M. Shayegan in Les Houches 1998,
``topological aspects of low-dimensional systems'' (Springer, NY, 2000),
A. Comtet, Th. Jolicoeur, S. Ouvry, F. David Eds.

\bibitem{Santos92}
M. B. Santos, Y. W. Suen, M. Shayegan, Y. P. Li, L. W. Engel, and D. C. Tsui,
Phys. Rev. Lett. \textbf{68}, 1188 (1992).

\bibitem{Santos92II}
M. B. Santos, J. Jo, Y. W. Suen, L. W. Engel, and M. Shayegan,
Phys. Rev. B\textbf{46}, 13639(R)  (1992).

%
%

\bibitem{Manoharan96}
H. C. Manoharan, Y. W. Suen, M. B. Santos, and M. Shayegan,
Phys. Rev. Lett. 77, 1813 (1996).

\bibitem{MengMa2020}
M. K. Ma, K. A. Villegas Rosales, H. Deng, Y. J. Chung, L. N. Pfeiffer, K. W. West, K. W. Baldwin,
R. Winckler, an M. Shayegan,
\phrl{125}{036601}{2020}.

\bibitem{HRY1999}
 E. H. Rezayi, F. D. M. Haldane, and K. Yang,
Phys. Rev. Lett. \textbf{83}, 1219 (1999).


\bibitem{HRY2000}
F. D. M. Haldane, E. H. Rezayi, and K. Yang,
Phys. Rev. Lett. \textbf{85}, 5396 (2000).

\bibitem{YHR2001}
K. Yang, F. D. M. Haldane, and E. H. Rezayi,
Phys. Rev. B\textbf{64}, 081301(R) (2001).

\bibitem{Wang2008}
H. Wang, D. N. Sheng, L. Sheng, and F. D. M. Haldane,
Phys. Rev. Lett. \textbf{100}, 116802 (2008).

\bibitem{McCann2012}
E. McCann and V. I. Fal'ko,
\phrl{96}{086805}{2006}.

\bibitem{Snizhko2012}
K. Snizhko, V. Cheianov, and S. H. Simon,
\phrb{85}{201415(R)}{2012}.

\bibitem{PapicAbanin}
Z. Papi\'c and D. A. Abanin,
Phys. Rev. Lett. \textbf{112}, 046602 (2014).

\bibitem{JTS}
Th. Jolicoeur, C. T\H{o}ke, and I. Sodemann,
Phys. Rev. B\textbf{99}, 115139 (2019).


\bibitem{JainBook}
J. K. Jain, ``Composite Fermions'' (Cambridge University
Press, Cambridge, England, 2007).

\bibitem{Haldane87}
F. D. M. Haldane,
in ``The Quantum Hall Effect'' by R. E. Prange and S. M. Girvin
(New York, Springer).

\bibitem{Papic2011}
Z. Papi\'c, D. A. Abanin, Y. Barlas, and R. N. Bhatt,
Phys. Rev. B\textbf{84}, 241306(R) (2011).

\bibitem{Sodemann}
Z. Zhu, D. N. Sheng, and I. Sodemann,
Phys. Rev. Lett. \textbf{124}, 097604 (2020).

\bibitem{Zhao}
Jianyun Zhao, Yuhe Zhang, and J. K. Jain,
\phrl{121}{116802}{2018}

\end{thebibliography}
\end{document}